\documentclass[document,12pt]{JHEP3}
\usepackage{latexsym}
\usepackage{graphics}
\usepackage{graphicx}
\usepackage{epsfig}
\usepackage{feynmp}
\usepackage{amssymb}
\newcommand{\mathsym}[1]{}

\newcommand{\be}{\begin{equation}}
\newcommand{\ee}{\end{equation}}
\newcommand{\bea}{\begin{eqnarray}}
\newcommand{\eea}{\end{eqnarray}}

\newcommand{\fr}{\frac}
\newcommand{\pt}{\partial}

\newcommand{\lt}{\left}
\newcommand{\rt}{\right}

\def\sp{\;\;\;,\;\;\;}
\def\hri#1#2{\href{http://arxiv.org/abs/#1}{[ArXiv:#1]#2}}
\def\hre#1#2{\href{http://arxiv.org/abs/#1/#2}{[ArXiv:#1/#2]}}

\vspace{-1cm}
\title{Heavy quarks  in a magnetic field}

\author{\href{http://hep.physics.uoc.gr/~kiritsis/}{Elias Kiritsis}$^{a,b}$, George Pavlopoulos$^a$\\
~\\
$^a$ \href{http://hep.physics.uoc.gr}{Crete Center for Theoretical Physics},
Department of Physics, University of Crete, 71003 Heraklion, Greece\\
~\\
$^b$ \href{http://www.apc.univ-paris7.fr}{APC, Universit\'e Paris 7}, \\ B\^atiment Condorcet, F-75205, Paris Cedex 13, France (UMR du CNRS 7164).}

\preprint{CCTP-2011-36}
\abstract{The motion of a heavy charged quark in a magnetic field is analyzed in the vacuum of strongly coupled CFT.
The motion of the quark is dissipative. It moves in spiral until it eventually comes to rest.
The world-sheet geometry is locally AdS$_2$ but has a time dependent horizon. The string profile in the static gauge extends from the boundary till a point where an embedding singularity exists. Connections with other circular string motions are established. }
\begin{document}
\maketitle

\section{Introduction}

\indent

The AdS/CFT correspondence provides us with a calculational tool
for large-$N_c$ gauge theories at strong coupling \cite{Maldacena}.
An interesting aspect of the correspondence is the duality between a heavy quark in gauge theory and a moving fundamental string end-point in the dual string theory.

In particular a heavy quark moving in the vacuum of N=4 sYM, would correspond to a string end-point, attached to a flavor brane at some radial position near the boundary and moving in AdS space.
Alternatively, if the quark moves in a plasma of temperature $T$, then the geometric background is replaced by the AdS-Schwarzschild black hole, with Hawking temperature T.

 Several cases of quarks moving in AdS thermal and non-thermal backgrounds have been analyzed so far, in \cite{Mikhailov's}-\cite{ChernicoffParedes}, developments that have been reviewed in \cite{gubserrev}. Quark energy loss in non-conformal backgrounds has been analyzed in \cite{transport}-\cite{langevin-2}.

 An interesting case that we will address in this paper is a heavy charged quark moving in a constant magnetic field. The charge can be a flavor charge (electric charge is a special case of this), and the magnetic field should be thought as being imposed on the flavor brane. Here we will impose the magnetic field at the endpoint of the string.
 The interest in this configuration stems from various contexts. Magnetic fields induce the chiral magnetic effect in strongly coupled matter, and this may have implications both for heavy ion experiments as well as neutron stars, \cite{Fukushima}.
 Magnetic fields are also one of the most important environments in condensed matter experiments. In view of the potential applications of holography to strongly coupled condensed matter systems, it is interesting to understand the physics of heavy colored objects in magnetic fields.
 The Hall conductivity, one of the main observables in this context, has been calculated in \cite{O'Bannon} and it would be interesting to eventually have a constituent understanding of the effects, as in the case of Ohmic conductivity, \cite{Karch}.

In this paper we will confine ourselves to AdS. We will use the Mikhailov solutions \cite{Mikhailov's}  in order to study heavy quark motion in a magnetic field and use the equations of motion derived in \cite{ChernicoffGuijosa}.
We will analyse the problem analytically, using perturbation theory for small motions and velocities, as well as numerically for arbitrary motion.
We will find that like a point particle in a magnetic field, the
motion is dissipative here. At the shifted boundary (flavor brane position), the quark end-point, instead of a circle, moves in a spiral that quickly collapses to a point.
In the dual gravitational description the energy-loss is happening through the flow energy through the gluonic string attached to the quark.

As usual energy is flowing down the string that is simulating the effects of a gluon cloud surrounding the quark and the energy radiated to infinity by the quark. The induced world-volume metric has a black-hole horizon, that is not static, but is moving towards the center of AdS.
Although this is not a purely thermal ensemble, the presence of a non-static horizon suggests a thermal nature, that can be made precise in the adiabatic limit.
Again dissipation and thermalization go hand in hand, as in previous related examples.

A related circular motion was studied in \cite{MITgroup}. In that case the world-sheet horizon position is static, and the end-point is making a constant circle at the boundary. This solution is different from ours and corresponds to a carefully tuned pair of electric and magnetic fields, pulsating appropriately.

We may use this solution to estimate the Hall conductivity of such carriers (at zero density and temperature) obtaining the classic Hall result.

At the end we also discuss the electromagnetic fields needed in order to have a circular motion of the string with constant angular momentum as in \cite{MITgroup}.

\indent
Our conclusions are as follows:
\begin{enumerate}
	\item The motion for the quark is a damped spiral, as the radiation emitted absorbs continuously its energy.
The velocity of the particle for late times is exponentially damped.

	\item There is a maximum initial velocity of the particle, beyond which the classical string description breaks down.

 \item The induced string metric has a horizon that is time dependent and its position moves exponentially fast at late times towards the center of AdS. The induced string metric is locally that of $AdS_2$ like in any other string motion in bulk AdS.

	\item In order to keep the trajectory of the particle to a fixed circle as in the MIT solution \cite{MITgroup}, we need a constant magnetic field and a time dependent electric field. The electric field must be proportional to the (rotating) velocity of the particle. In this way the electric field provides energy to the endpoint, equal to the energy radiated to infinity by the gluonic field.

	\item The radiation emitted by the quark is exponentially damped with time for large times.

	\item The embedding $X^{\mu}(\tau,r)$ of the string with respect to the proper time $\tau$ on the boundary and the radial direction $r$ of the AdS were found in \cite{ChernicoffGuijosa}.
	The embedding of the string in the static gauge $\vec{X}(X^0,r)$ stops at a point where it has an embedding singularity. In some of the cases that we managed to check numerically, the embedding singularity is hidden by the world-sheet horizon.

\end{enumerate}

\section{General setup}

\indent

We consider a string moving in pure 5d AdS spacetime in the Poincare patch with metric
\bea
ds^2=\fr{L^2}{r^2}\lt(-dt^2+dr^2+dx^2+dy^2+dz^2\rt)\;.
\eea
One endpoint of the string lies on a flavour brane at $r=\Lambda$ on which there is a constant magnetic field $\vec{B}$ in a spatial 3d subspace of the brane.\\
We use the Mikhailov solution \cite{Mikhailov's} for the motion of the string in pure AdS and apply the boundary conditions on the endpoint on the flavour brane.\\
We consider the endpoint to represent the motion of a massive quark. The energy flow from the endpoint down the string represents
 the radiation emitted (strongly coupled gluons) on the flavour brane through the AdS/CFT correspondence.\\
We also consider the Hall effect by adding a small electric field perpendicular to the magnetic one and examine the motion of the electric charge.

\section{Motion of a string in pure AdS}

We consider the Nambu-Goto action for the string in the bulk of AdS
and the electromagnetic coupling at the endpoint of the string on a flavour brane at $r=\Lambda$:
\bea
S=S_{NG}+S_{F}\sp S_{NG}=-\fr{1}{2\pi \ell^2_s}\int{d^2\sigma\sqrt{-detg_{ab}}}\sp S_{F}=\int{d\tau A_{\mu}}\lt(x\lt(\tau\rt)\rt)\pt_{\tau}x^{\mu} \nonumber \\
\eea
in terms of the quark world-line at $r=\Lambda=\fr{\sqrt{\lambda}}{2\pi m}$ where
 $m$ is the mass of the quark and $\lambda=\fr{L^4}{\ell_s^4}=g_{YM}^2N_c,$ the
  't Hooft coupling\footnote{We name by $L$ the size of the AdS throat, $\ell_s$
  the length of the string, $g_{YM}$ the coupling constant of the gauge theory on the boundary and $N_c$ the number of colors.}.We name by $x^\mu$ the coordinates and by $\tau$ the proper time of the endpoint of the string on the boundary at $r=\Lambda$ and by $A^\mu$ the gauge field on the same boundary.

The boundary conditions at $r=\Lambda$ are
\bea
\Pi_{\mu}^{r}\lt(\tau\rt)|_{r=\Lambda}=\mathcal{F}_{\mu}(\tau)\quad \forall\ \tau
\eea
where $\mathcal{F}^\mu$ is the Lorentz four-force exerted on the endpoint on the boundary and
\bea
\Pi_{\mu}^{r}\equiv\fr{\pt \mathcal{L}_{NG}}{\pt\lt(\pt_{r}X^{\mu}\rt)}=\fr{\sqrt{\lambda}}{2\pi}\lt(\fr{\lt(\pt_{\tau}X\rt)^2
\pt_{r}X^{\mu}-\lt(\pt_{\tau}X \cdot \pt_{r}X\rt)\pt_{\tau}X^{\mu}}{r^2\sqrt{\lt(\pt_{\tau}X \cdot \pt_{r}X\rt)^2
-\lt(\pt_{\tau}X\rt)^2\lt(1+\pt_{r}X^2\rt)}}\rt)
\eea
is the Nambu-Goto boundary term. \\
The Lorentz four-force exerted on the quark on the boundary is
\bea
\mathcal{F}_{\mu}=-F_{\mu\nu}\pt_{\tau}x^{\nu}.
\eea

\paragraph{Equation of motion  for the quark}

In the case our quark has finite mass $m,$ we consider the endpoint of the quark to be at finite $\Lambda.$
We are using the retarded solution of Mikhailov \cite{Mikhailov's} for the motion of the string
\bea
X^{\mu}(\tilde{\tau},r)=\tilde{x}^{\mu}(\tilde{\tau})+r\fr{d\tilde{x}^{\mu}}{d\tilde{\tau}}.
\label{mikhsol}
\eea

The tilded coordinates refer to an auxiliary boundary at $r=0$ we consider because for these coordinates the solution of Mikhailov is much simpler.
Then, $\tilde{x}^{\mu}$ are the coordinates $\tilde{t},\tilde{x},\tilde{y},\tilde{z}$ of the auxiliary endpoint at $r=0$ and $\tilde{\tau}$ its proper time.
With respect to the coordinates of the quark at $r=\Lambda$ the Mikhailov solution can be written as \cite{ChernicoffGuijosa}:
\bea
X^\mu(\tau,r)=\lt(\fr{r-\Lambda}{\sqrt{1-\Lambda^4\fr{4\pi^2}{\lambda}\mathcal{F}^2}}\rt)\lt(\fr{dx^\mu}{d\tau}-\fr{2\pi}{\sqrt{\lambda}}\Lambda^2\mathcal{F}^\mu\rt)+x^\mu(\tau).
\label{stringsol}
\eea
where $\tau$ the proper time of the quark,  $x^{\mu}=\{t,x,y,z\}$ the space time coordinates of the quark, $\mathcal{F}^{\mu}$ the 4-force on the quark and $m$ its mass.

The differential equation describing the motion of the quark is
\bea
\fr{d}{d\tau}\lt(\fr{m\fr{dx^\mu}{d\tau}-\fr{\sqrt{\lambda}}{2\pi m}\mathcal{F}^\mu}{\sqrt{1-\fr{\lambda}{4\pi^2m^4}\mathcal{F}^2}}\rt)=\fr{\mathcal{F}^\mu-\fr{\sqrt{\lambda}}{2\pi m^2}\mathcal{F}^2\fr{dx^{\mu}}{d\tau}}{1-\fr{\lambda}{4\pi^2 m^4}\mathcal{F}^2}.
\eea
This differential equation with the substitution $\mathcal{F}_{\mu}=-F_{\mu\nu}\pt_{\tau}x^{\nu}$ for the force due to the electromagnetic fields gives us the motion of the endpoint,i.e. the functions $x(\tau),y(\tau).$\\
Then, with the help of these functions for the endpoint, we can find the motion of the whole string using (\ref{stringsol}).\\
For the reasons discussed in the introduction, we will examine the motion of the quark in a constant magnetic field.

\section{The case of constant magnetic field on the boundary}

 We will consider the case where the magnetic field $\vec{B}$ at the boundary, $r=\Lambda$ is constant.
For convenience we consider it to be oriented towards the $z$ axis, namely $\vec{B}=B\hat{z}.$ The four-force on the quark becomes
\bea
\mathcal{F}^0=0\sp\mathcal{F}^{x}(\tau)=-B\dot{y}(\tau)\sp\mathcal{F}^{y}(\tau)=B\dot{x}(\tau)
\eea
and the equations of motion are
\bea
\fr{d}{d\tau}\lt(\fr{m\fr{dt}{d\tau}}{\sqrt{1-\fr{\lambda}{4\pi^2m^4}B^2\lt(\fr{d\vec{x}}{d\tau}\rt)^2}}\rt)&=&-\fr{\fr{\sqrt{\lambda}}{2\pi m^2}B^2\lt(\fr{d\vec{x}}{d\tau}\rt)^2\fr{dt}{d\tau}}{1-\fr{\lambda}{4\pi^2 m^4}B^2\lt(\fr{d\vec{x}}{d\tau}\rt)^2}. \label{eomt}\\
\fr{d}{d\tau}\lt(\fr{m\fr{d\vec{x}}{d\tau}-\fr{d\vec{x}}{d\tau}\times\vec{B}}{\sqrt{1-\fr{\lambda}{4\pi^2m^4}B^2\lt(\fr{d\vec{x}}{d\tau}\rt)^2}}\rt)&=&\fr{\fr{d\vec{x}}{d\tau}\times\vec{B}-\fr{\sqrt{\lambda}}{2\pi m^2}B^2\lt(\fr{d\vec{x}}{d\tau}\rt)^2\fr{d\vec{x}}{d\tau}}{1-\fr{\lambda}{4\pi^2 m^4}B^2\lt(\fr{d\vec{x}}{d\tau}\rt)^2}.
\label{eoms}
\eea

Equation (\ref{eomt}) can be derived from (\ref{eoms}) using
\be
d\tau^2=-dx^{\mu}dx_{\mu}\Rightarrow d\tau^2=dt^2-d\vec{x}^2.
\ee
We also define the dimensionless constant $s:$
\be
B=s\fr{m}{\Lambda}.
\label{sdef1}
\ee
We choose units so that  $\Lambda=\fr{\sqrt{\lambda}}{2\pi m}=1\Rightarrow m=\fr{\sqrt{\lambda}}{2\pi},$ i.e. we measure $x(\tau),y(\tau),\tau,t$ in units of $\Lambda$.
 Then (\ref{sdef1}) becomes
\be
B=s\ m.
\ee
The equations of motion (\ref{eoms}) can be rewritten as
\bea
\fr{d}{d\tau}\lt(\fr{\fr{dt}{d\tau}}{\sqrt{1-s^2\lt(\fr{d\vec{x}}{d\tau}\rt)^2}}\rt)&=&-\fr{s^2\lt(\fr{d\vec{x}}{d\tau}\rt)^2\fr{dt}{d\tau}}{1-s^2\lt(\fr{d\vec{x}}{d\tau}\rt)^2}\label{eoms1} \\
\fr{d}{d\tau}\lt(\fr{\fr{d\vec{x}}{d\tau}-s\fr{d\vec{x}}{d\tau}\times\hat{z}}{\sqrt{1-s^2\lt(\fr{d\vec{x}}{d\tau}\rt)^2}}\rt)&=&\fr{s\fr{d\vec{x}}{d\tau}\times\hat{z}-s^2\lt(\fr{d\vec{x}}{d\tau}\rt)^2\fr{d\vec{x}}{d\tau}}{1-s^2\lt(\fr{d\vec{x}}{d\tau}\rt)^2}.
\label{eoms2}
\eea
Because $\fr{dt}{d\tau}=\fr{1}{\sqrt{1-\lt(\fr{d\vec{x}}{dt}\rt)^2}}\equiv\gamma,$ the first equation actually follows from the two others.
Therefore, we shall find the functions $x(t),y(t)$ by solving the differential equations (\ref{eoms2}).

 Using $\fr{dt}{d\tau}=\fr{1}{\sqrt{1-\lt(\fr{d\vec{x}}{dt}\rt)^2}}$ we obtain the following equations for the spatial transverse coordinates $x(t),y(t):$
\bea
\fr{d}{dt}\lt(\gamma\fr{\fr{d\vec{x}}{dt}-s\fr{d\vec{x}}{dt}\times\hat{z}}{\sqrt{1-s^2\gamma^2\lt(\fr{d\vec{x}}{dt}\rt)^2}}\rt)&=&\fr{s\fr{d\vec{x}}{dt}\times\hat{z}-s^2\gamma^2\lt(\fr{d\vec{x}}{dt}\rt)^2\fr{d\vec{x}}{dt}}{1-s^2\gamma^2\lt(\fr{d\vec{x}}{dt}\rt)^2}.
\label{eomst}
\eea
where $\gamma=\fr{dt}{d\tau}=\fr{1}{\sqrt{1-\lt(\fr{d\vec{x}}{dt}\rt)^2}}$ is the dilatation factor.

Solving (\ref{eomst}) for $\ddot{\vec{x}}$ we obtain:
\bea
\ddot{\vec{x}}=-\fr{s\sqrt{1-(\gamma^2-1)s^2}}{\gamma(1+s^2)}\lt(s\dot{\vec{x}}+\hat{z}\times\dot{\vec{x}}\rt).
\label{eomsddx}
\eea
We observe that there is a maximum velocity $v_{max}$ for the quark when the subroot quantity in (\ref{eomsddx}) becomes zero.
The maximum velocity $v_{max}$ is given by:
\bea
1-(\gamma^2-1)s^2=1-\fr{s^2v_{max}^2}{1-v_{max}^2}=0\Rightarrow v_{max}=\fr{1}{\sqrt{1+s^2}}.
\label{vmax}
\eea
Therefore, in the presence of a constant magnetic field on the boundary, the initial velocity of the quark in the beginning of the motion must be smaller than $v_{max}$.
For small $s<<1$,  the initial velocity can be in the relativistic regime, whereas for large $s>>1$ the initial velocity of the motion can be only non relativistic.

\indent

When the particle moves with velocity near $v_{max}$, this corresponds to velocity near to that of light on the auxiliary boundary at $r=0.$
This can be easily seen by \cite{ChernicoffGuijosa}
\bea
\fr{dx^\mu}{d\tilde{\tau}}&=&\fr{d^2\tilde{x}^\mu}{d\tilde{\tau}^2}+\fr{d\tilde{x}^\mu}{d\tilde{\tau}},\fr{d\tilde{\tau}}{d\tau}=\fr{1}{\sqrt{1-s^2\gamma^2\vec{v}^2}}\Rightarrow\nonumber \\
\fr{d\tau}{d\tilde{\tau}}\gamma \vec{v}&=&\fr{1}{\sqrt{1-s^2\gamma^2\vec{v}^2}}(s\vec{v}\times\hat{z}-s^2\gamma^2\vec{v}^2\ \vec{v})+\tilde{\gamma}\vec{\tilde{v}}\Rightarrow\nonumber \\
\sqrt{1-s^2\gamma^2\vec{v}^2}\gamma \vec{v}&=&\fr{1}{\sqrt{1-s^2\gamma^2\vec{v}^2}}(s\vec{v}\times\hat{z}-s^2\gamma^2\vec{v}^2\ \vec{v})+\tilde{\gamma}\vec{\tilde{v}}
\label{vr0c}
\eea
where by $\tilde{\tau},\tilde{x}^\mu$ we name the proper time and the coordinates on the auxiliary boundary at $r=0,$
and by $\vec{\tilde{v}},\tilde{\gamma}$ we name the velocity and gamma factor at the auxiliary boundary at $r=0.$
The corresponding untilded quantities correspond to the true boundary at $r=\Lambda=1.$
We observe in (\ref{vr0c}) that for $||\vec{v}||\rightarrow v_{max},$ the l.h.s. tends to $0,$ whereas the term $\fr{1}{\sqrt{1-s^2\gamma^2\vec{v}^2}}(s\vec{v}\times\hat{z}-s^2\gamma^2\vec{v}^2\ \vec{v})$ tends to $\infty.$ Therefore, the term $\tilde{\gamma}\vec{\tilde{v}}$ must be a vector with the same length (and opposite direction) with the first term of the r.h.s. and this means $\tilde{\gamma}=\infty$ and therefore $||\vec{\tilde{v}}||= c.$

Multiplying (\ref{eomsddx}) with $\dot{\vec{x}}$ we obtain
\bea
\ddot{\vec{x}} \cdot \dot{\vec{x}}= \fr{1}{2}\fr{d\lt(\dot{\vec{x}}^2\rt)}{dt}=-\fr{s^2\sqrt{1+(1-\gamma^2)s^2}}{\gamma(1+s^2)}\dot{\vec{x}}^2
\label{v2t}
\eea
that suggests that the motion is damped and after some time the quark will practically stop moving.
Then we can integrate (\ref{v2t}) and obtain the function $t(\vec{v}^2)$
\bea
\int_{0}^{t}{dt}&=&-\int_{v_0^2}^{v(t)^2}\fr{1}{2\fr{s^2\sqrt{1+(1-\gamma^2)s^2}}{\gamma(1+s^2)}v^2}d\lt(v^2\rt)  \Rightarrow \nonumber \\
t(v^2)&=&\frac{\left(s^2+1\right) \lt(\tanh ^{-1}\left(\sqrt{1-\left(s^2+1\right) v^2}\right)-\tanh^{-1}\left(\sqrt{1-\left(s^2+1\right) v_0^2}\right)\rt)}{s^2}\nonumber  \\
\eea
which gives for $v(t)^2$:
\bea
v(t)^2=\frac{sech^2\left(\frac{s^2 t}{s^2+1}+\tanh ^{-1}\left(\sqrt{1-\left(s^2+1\right)
   v_0^2}\right)\right)}{s^2+1}.
\label{vsqt}
\eea

We multiply externally $\dot{\vec{x}}$ with (\ref{eomsddx}), we take the component in the $z-$direction\footnote{We remind that we have named with $r$ the radial direction of AdS with metric $$ds^2=-\fr{L^2}{r^2}\lt(-dt^2+dr^2+dx^2+dy^2+dz^2\rt)$$ and with $x,y,z,t$ the tranverse directions.} and we obtain
\bea
\lt(\dot{\vec{x}}\times\ddot{\vec{x}}\rt)_z&=&-\fr{s\sqrt{1+(1-\gamma^2)s^2}}{\gamma(1+s^2)}\lt(\dot{\vec{x}}\times
\lt(\hat{z}\times\dot{\vec{x}}\rt)\rt)_z=-\fr{s\sqrt{1+(1-\gamma^2)s^2}}{\gamma(1+s^2)}\dot{\vec{x}}^2\Rightarrow\nonumber\\
||\dot{\vec{x}}|| ||\ddot{\vec{x}}||\sin\lt(\theta_{\vec{v},\vec{a}}\rt)&=&-\fr{s\sqrt{1+(1-\gamma^2)s^2}}{\gamma(1+s^2)}\dot{\vec{x}}^2,
\label{vaext}
\eea
where we have named with $\theta_{\vec{v},\vec{a}}$ the angle from $\vec{v}$ to $\vec{a}$.\footnote{This angle can be positive and negative and lies in the range $(-\pi,\pi]$ because the cross product $\vec{v}\times\vec{a}$ is sensitive to which vector we choose first and which second.}

We can divide $\lt(\dot{\vec{x}}\times\ddot{\vec{x}}\rt)_z$ by $\fr{1}{2}\fr{d\lt(v^2\rt)}{dt}=||\vec{v}|| ||\vec{a}|| \cos\lt(\theta_{\vec{v},\vec{a}}\rt)$ in (\ref{v2t}) and we obtain
\be
\fr{\left(\vec{v}\times\vec{a}\right)_z}{\vec{v}\cdot\vec{a}}=\fr{1}{s}.
\ee

We note from (\ref{vaext}) that for positive $s$ ($B>0$ codirectional with $\hat{z}$) the angle from $\vec{v}$ to $\vec{a}$ is negative, whereas for negative $s,$ the angle from $\vec{v}$ to $\vec{a}$ is positive. Therefore, the tangent of the angle $|\theta_{\vec{v},\vec{a}}|$ between the vectors $\vec{v},\vec{a}$ measured in the range $0,\pi$ is

\be
\tan\lt(|\theta_{\vec{v},\vec{a}}|\rt)=-\fr{1}{s}\;.
\ee
It is negative as expected for deceleration and it is a constant during the motion.

\subsection{Energy carried by the quark and radiation emission}

\indent

As mentioned in \cite{ChernicoffGuijosa} the four-momentum of the quark
and the rate at which four-momentum is carried away are correspondingly
\bea
p_{q}^\mu=\fr{m\fr{dx^{\mu}}{d\tau}-\fr{\sqrt{\lambda}}{2\pi m}\mathcal{F}^{\mu}}{\sqrt{1-\fr{\lambda}{4\pi^2m^4}\mathcal{F}^2}}
\sp \fr{dP_{rad}^{\mu}}{d\tau}=\fr{\sqrt{\lambda}\mathcal{F}^2}{2\pi m^2}\lt(\fr{\fr{dx^{\mu}}{d\tau}-\fr{\sqrt{\lambda}}{2\pi m^2}\mathcal{F}^{\mu}}{1-\fr{\lambda}{4\pi^2m^4}\mathcal{F}^2}\rt),
\label{419}\eea
where by $p_q$ we note the energy of the dressed quark (energy of the quark$+$near gluonic field) and by $\fr{dP_{rad}}{d\tau}$ we name the rate at which energy is radiated towards infinity by the quark.
As mentioned in \cite{ChernicoffGuijosa} in the case of a heavy quark ($\fr{\sqrt{\lambda|\mathcal{F}^2|}}{2\pi m^2}=\fr{s\ B||\vec{v}|| \gamma}{m}<<1$) the equation of motion gives
\bea
m \lt(\fr{d^2x^\mu}{d\tau^2}-\fr{\sqrt{\lambda}}{2\pi m}\fr{d^3x^\mu}{d\tau^3}\rt)=\mathcal{F}^\mu-\fr{\sqrt{\lambda}}{2\pi}\fr{d^2x^\nu}{d\tau^2}\fr{d^2x_\nu}{d\tau^2}\fr{dx^\mu}{d\tau},
\eea
and on the r.h.s. we recognize the radiation reaction force given by the covariant Lienard formula, as expected from \cite{Mikhailov's}
for the case of the endpoint on the boundary of AdS at $r=0$.

In our case (\ref{419}) can be written as
\be
\fr{dE_{rad}}{dt}=\frac{m s^2 \dot{\vec{x}}(t)^2}{\sqrt{1-\dot{\vec{x}}(t)^2} \left(1-\left(s^2+1\right) \dot{\vec{x}}(t)^2\right)} \sp
\fr{d\vec{P}_{rad}}{dt}=\lt(\dot{\vec{x}}(t)+s\dot{\vec{x}}(t)\times\hat{z}\rt)\fr{dE_{rad}}{dt}
\label{derad}
\ee

We note that the rate of energy transfer $\fr{dE_{rad}}{dt}$ is an increasing function of $||\dot{\vec{x}}||$
which means that the faster the particle moves, the faster is the rate at which it loses its energy.\\
By using the expression (\ref{vsqt}) for the velocity as a function of time we obtain
\bea
\fr{dE_{rad}}{dt}=\frac{m s^2 csch^2\left(\frac{s^2 t}{s^2+1}+\tanh ^{-1}\left(\sqrt{1-\left(s^2+1\right)
   v_0^2}\right)\right)}{\sqrt{\left(s^2+1\right) \left(sech^2\left(\frac{s^2 t}{s^2+1}+\tanh
   ^{-1}\left(\sqrt{1-\left(s^2+1\right) v_0^2}\right)\right)+s^2+1\right)}}.
\eea
We have drawn the plots of the rate of energy loss per unit mass for a particle with initial velocity $v_0$ near $v_{max}$ (\ref{vmax}), for small parameter $s$ in the right figure \ref{fig:velrelss} and for large $s$ in the right figure \ref{fig:velrells}. We observe that for late times $t>>\beta^{-1}=\fr{s^2+1}{s^2}$
\bea
\fr{dE_{rad}}{dt}\propto e^{-\fr{2s^2 t}{1+s^2}},
\eea
has an exponential damping as the squared velocity $v(t)^2.$

\subsection{The linear approximation \label{linearapp}}

The equations of motion derived above are non-linear. However in the non-relativistic regime they linearize.
In this case,  we can approximate the equations (\ref{eomsddx}) with the following linear equations,
\bea
\lt(
\begin{array}[h]{c}
\ddot{x}^{(0)} \\
\ddot{y}^{(0)}	
\end{array}
\rt)&=&\lt(
\begin{array}[h]{cc}
-\frac{s^2}{1+s^2} &\frac{s}{1+s^2} \\
-\frac{s}{1+s^2} &-\frac{s^2}{1+s^2}
\end{array}
\rt)
\lt(
\begin{array}[h]{c}
\dot{x}^{(0)} \\
\dot{y}^{(0)}	
\end{array}
\rt).
\eea
with solution
\bea
\lt(
\begin{array}[h]{c}
x^{(0)}(t)
\\	
y^{(0)}(t)
\end{array}
\rt)=R_0e^{-\fr{s^2}{1+s^2}t}\lt(
\begin{array}[h]{c}
\cos(\frac{s}{1+s^2} t+\phi_0)
\\	
\sin(\frac{s}{1+s^2} t+\phi_0)
\end{array}
\rt)+
\lt(
\begin{array}[h]{c}
A\\	
B
\end{array}
\rt).
\label{linsol}
\eea

The constants $\omega,\beta$ describe the angular velocity and the damping factor of the motion of the quark correspondingly.They are given by
\bea
\omega=\frac{s}{1+s^2}\sp \beta=\fr{s^2}{1+s^2}.
\label{omlam}
\eea

\indent

In order for the linear regime  (\ref{eomsddx}) to be reliable,  we must have $v_0<<v_{max}=\fr{1}{\sqrt{1+s^2}}$ for the initial velocity $v_0.$
For small $s$,  the condition $v_0^2<<1$ is sufficient, whereas for large $s$ the condition $v_0<<\fr{1}{s}$ is the relevant one.

\section{The nonlinear motion}

We now consider the full non-linear equations of motion. Their behavior depends importantly on the size of the external magnetic field and we will analyze the two cases separately below.

\begin{figure}
	\centering
		\epsfig{file=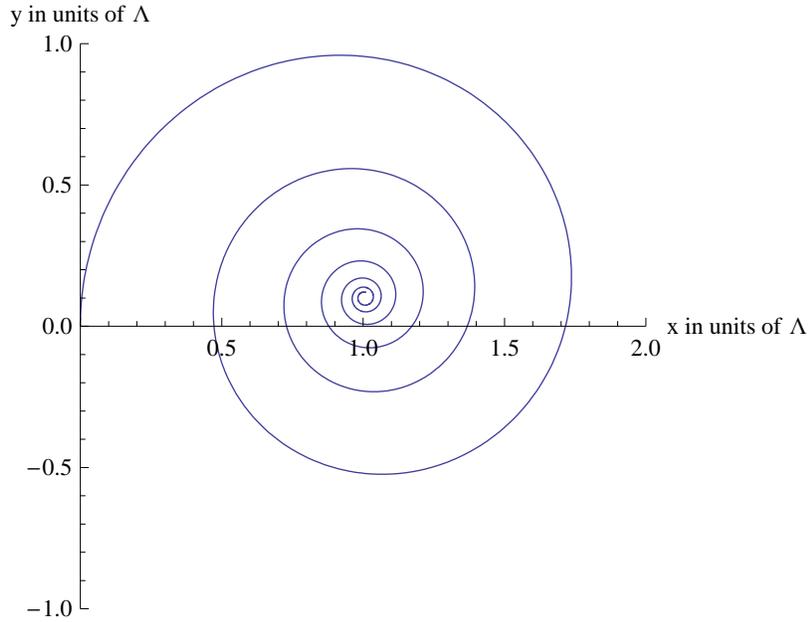}
		\caption{Trajectory of the particle for $v_0=0.1c,s=0.1$}
	\label{fig:graphnonrelss}
\end{figure}
\begin{figure}
	\centering
		 \epsfig{file=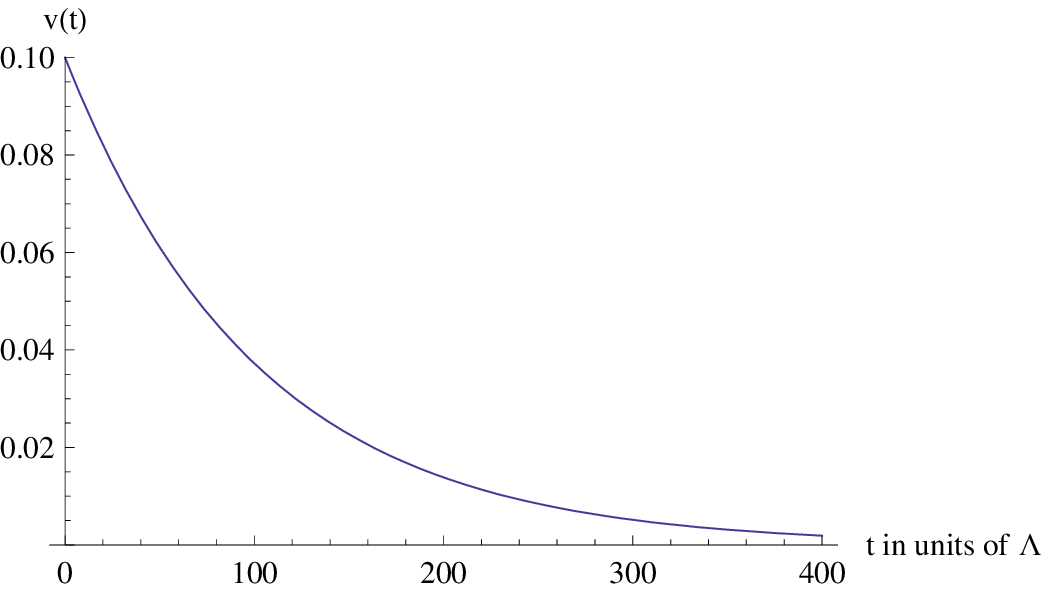,height=2in,width=3in}\epsfig{file=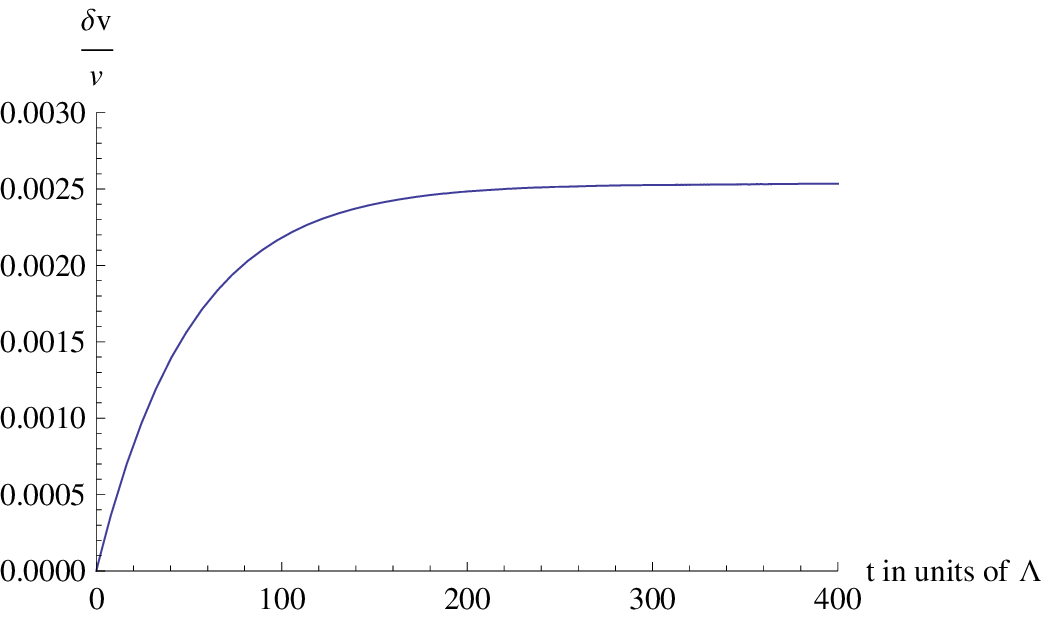,height=2in,width=3in}
		\caption{\textbf{Left}:Velocity of the particle for $v_0=0.1c,s=0.1$,\qquad
		\textbf{Right}:Relative difference $\fr{v_{num(t)-v_{lin.app.}(t)}}{v_{num}(t)}$ in the velocity of the particle for $v_0=0.1c,s=0.1$}
\label{fig:velnonrelss}
\end{figure}

\subsection{Weak magnetic field}

We first set $s\ll 1$ and as an illustrative example we use $s=0.1,$ which gives $v_{max}\approx0.995c$ for the maximum initial velocity the motion can start with.
We consider the quark to have initial velocity $v_0$ and we integrate numerically the motion of the quark under the nonlinear equations (\ref{eoms2}), which solved for $\ddot{\vec{x}}$ gives
\be
\ddot{\vec{x}}=-\fr{s\sqrt{1+(1-\gamma^2)s^2}}{\gamma(1+s^2)}\lt(s\dot{\vec{x}}-\dot{\vec{x}}\times\hat{z}\rt)\sp \gamma=\fr{1}{\sqrt{1-\lt(\fr{d\vec{\vec{x}}}{dt}\rt)^2}}
\ee
Various aspects of the motion are portrayed in the figures.

In figure \ref{fig:graphnonrelss} we show the trajectory of the quark on the $x-y$ plane at the shifted boundary $r=\Lambda$,  for a non-relativistic initial velocity $v_0=0.1c<<v_{max}\approx0.995c$ and small parameter $s=0.1$. Distances are in units of $\Lambda$.

In the left figure \ref{fig:velnonrelss} we show the velocity of the quark as a function of time $t$ .\footnote{We remind the reader that $r=\Lambda$ is the position of the flavour brane on which the string endpoint lives.}
In the right figure \ref{fig:velnonrelss} we show the relative difference in the velocity of the particle with time between the numerical solution and the linear approximation for the same initial velocity $v_0=0.1c$ and $s=0.1\ $. The corrections to the linearized equations are of order $v_0^2$ and $s^2v_0^2$ and for $s<<1$ the corrections of order $v_0^2$ are the relevant ones. Therefore we expect the relative difference of velocities between the linear approximation and the nonlinear motion to be of order $v_0^2\approx0.01=1\%$ in accordance with the observed $\fr{v_{num(t)-v_{lin.app.}(t)}}{v_{num}(t)}\rightarrow0.0025=0.25\%.$

\begin{figure}
	\centering
		\epsfig{file=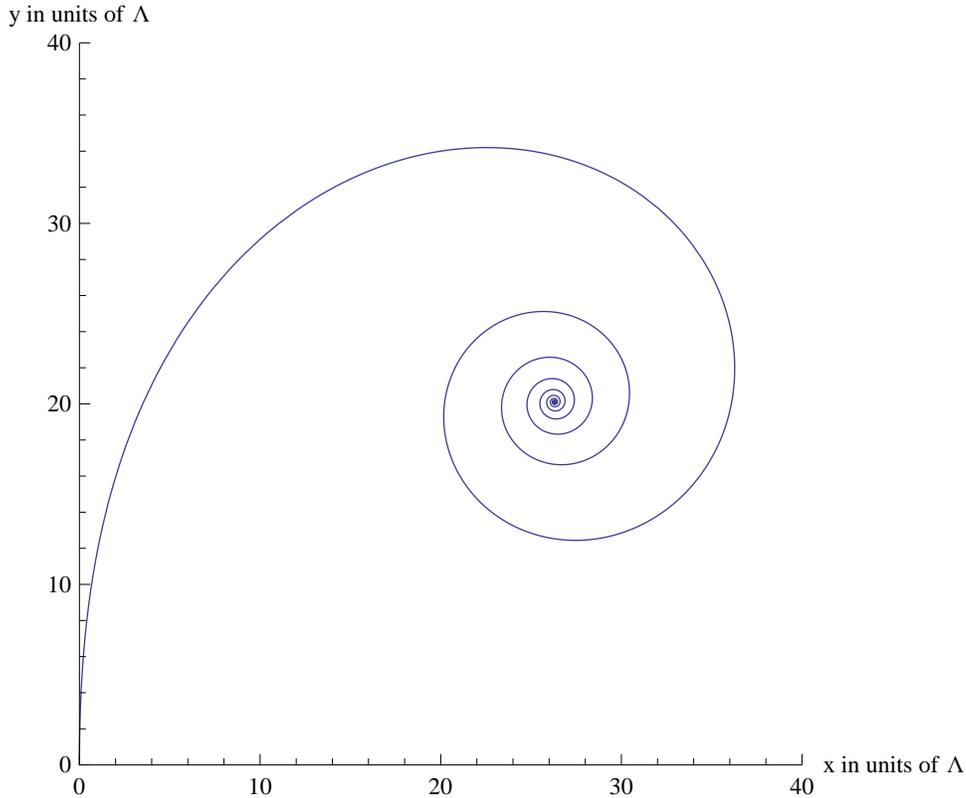}
		\caption{Trajectory of the particle for relativistic initial velocity $v_0=0.99c$ and small $s=0.1.$}
	\label{fig:graphrelss}
\end{figure}

\begin{figure}
	\centering
		 \epsfig{file=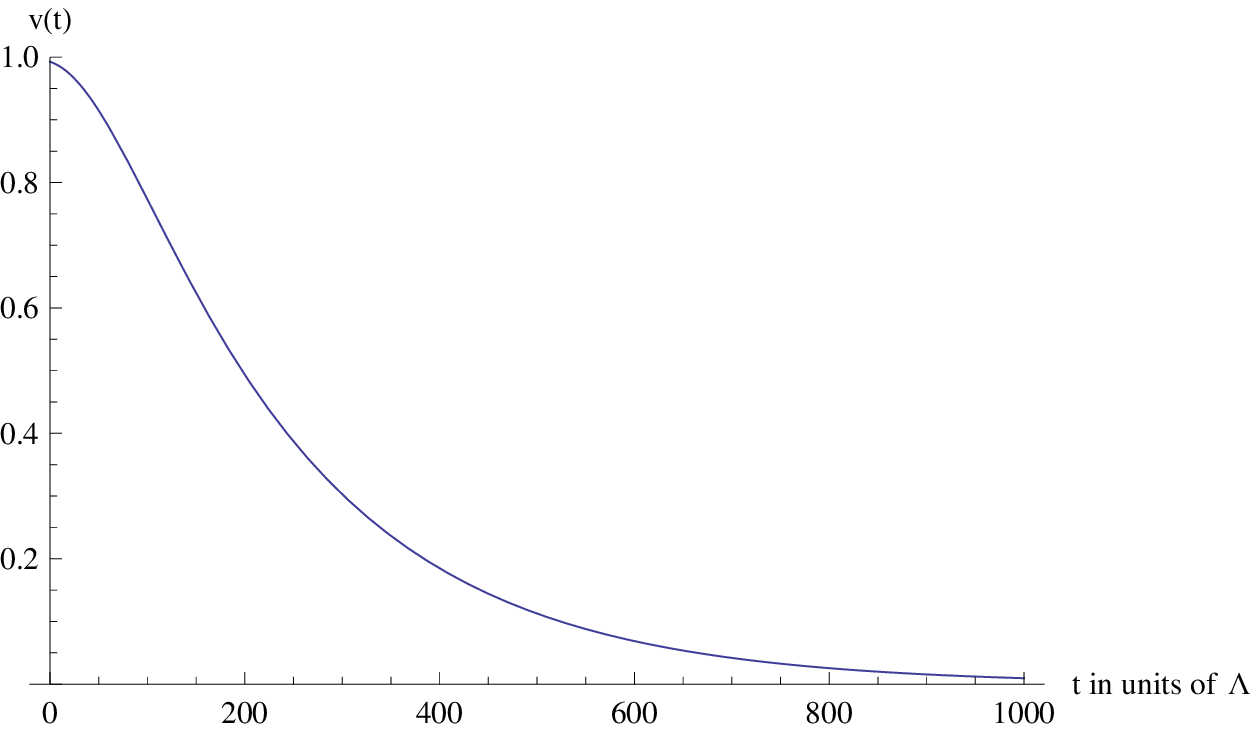,height=2in,width=3in}\epsfig{file=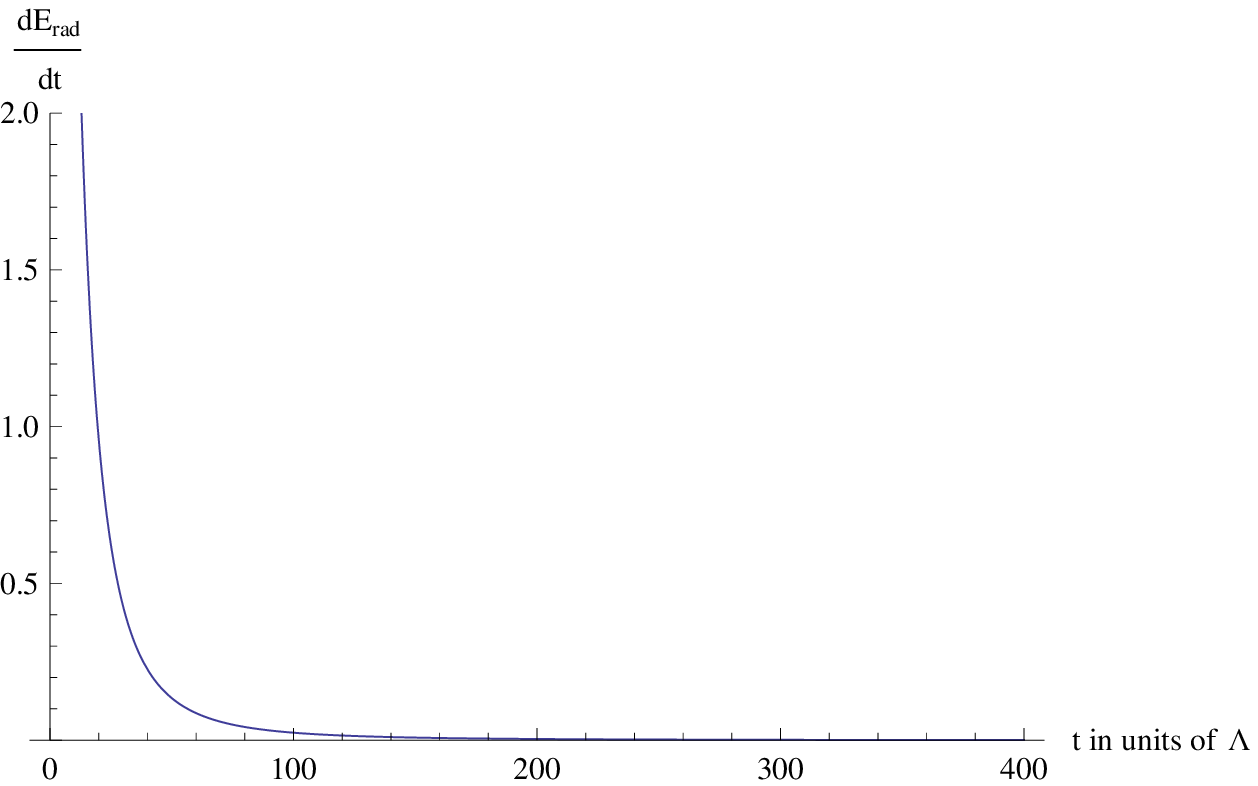,height=2in,width=3in}
		\caption{\textbf{Left}:Velocity of the particle for small $s=0.1$ and relativistic initial velocity $v_0=0.99c\approx v_{max}.$\qquad
\textbf{Right}:Rate of energy transfer per unit mass from the particle for small $s=0.1$ and initial velocity $v_0=0.99c\approx v_{max}.$ }	
	\label{fig:velrelss}
\end{figure}

In figure \ref{fig:graphrelss} we consider relativistic initial velocity $v_0=0.99c\approx v_{max}=0.995c$ and small magnetic field $s=0.1\ $. The trajectory of the endpoint is shown.
In the left figure \ref{fig:velrelss} we show the velocity of the particle as a function of time $t$.
In the right figure \ref{fig:velrelss} we show the rate of energy transfer per unit mass of the particle for small $s=0.1$ and relativistic initial velocity $v_0=0.99c\approx v_{max}$.

\subsection{Strong magnetic field}

In the case of a strong magnetic field ($s\gg 1$), the maximum initial velocity of the quark $v_0$ can be
\be
v_{max}=\fr{1}{\sqrt{1+s^2}}\approx\fr{1}{s}.
\ee
In order for the linear approximation to be valid we must have for the initial velocity
$v_{0}<<v_{max}\approx\fr{1}{s}$.

\begin{figure}
	\centering
		 \epsfig{file=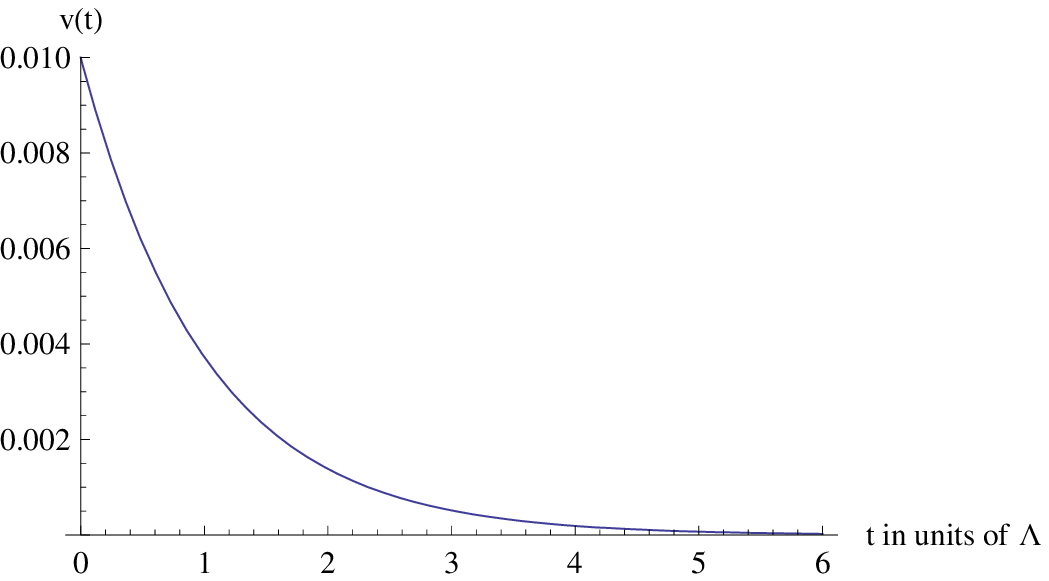,height=2in,width=3in}\epsfig{file=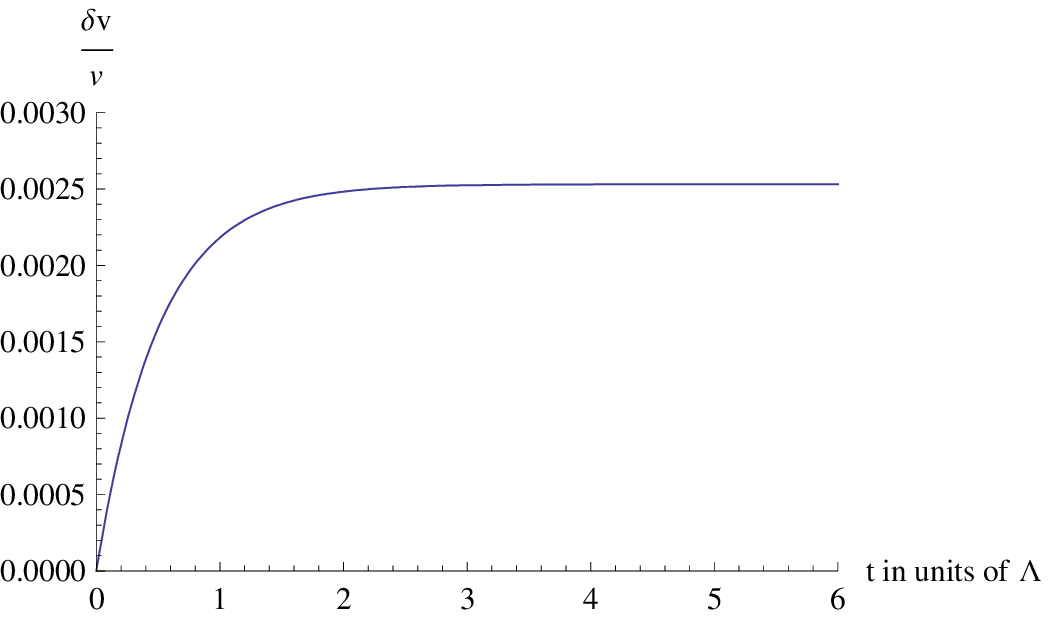,height=2in,width=3in}
		\caption{\textbf{Left}:Velocity of the particle for large $s=10$ and initial velocity $v_0=0.01c<<v_{max}\approx0.1c.$\qquad
		\textbf{Right}:Relative difference $\fr{v_{num(t)-v_{lin.app.}(t)}}{v_{num}(t)}$ in the velocity of the particle for $v_0=0.01c,s=10,s^2v_0^2=0.01<<1.$}
	\label{fig:velnonrells}
\end{figure}

In the left figure \ref{fig:velnonrells} we show the velocity of the quark as a function of time for large parameter $s=10$ and for initial velocity $v_0=0.01c<<v_{max}\approx0.1c.$
In this case the motion is in the validity region of the linear approximation. Therefore, we can compare the velocity calculated numerically with the velocity obtained from the linear approximation and their relative difference is shown in the right figure \ref{fig:velnonrells}.
Because the corrections to the linear equations are of order $v_0^2$ and $s^2v_0^2,$ for $s\gg 1$ the correction $s^2v_0^2$ is the relevant one and we expect the relative difference of the velocities to be of order $\fr{\delta v}{v}\approx s^2v_0^2\approx0.01=1\%$ for large $s.$ This is satisfied by the numerical solution which has $\fr{\delta v}{v}\rightarrow0.0025=0.25\%$.

In the left figure \ref{fig:velrells} we show the velocity of the quark as a function of time for initial velocity $v_0=0.099c$ and parameter $s=10.$ In this case,  $v_0$ is very close to the maximum allowed $v_{max}=\fr{1}{\sqrt{1+s^2}}\approx\fr{1}{s}$ and therefore the linear approximation is invalid. This is evident by the shape of the corresponding curve near $v_0$ which is not of exponential form.\\

In the right figure \ref{fig:velrells} we show the rate of energy transfer per unit mass of the particle for large $s=10$ and initial velocity near $v_{max}$.

\subsection{Comments on the nonlinear trajectories}

A generic feature of our solutions is that the motion is damped due to energy transfer from the endpoint to the string, which is equivalent in terms of the gauge theory on the $4d$ boundary to the transfer of energy from the quark to the gluonic degrees of freedom.
The timescale characteristic of the damping is $\beta^{-1}=\fr{1+s^2}{s^2}$ in the linearized regime.
This is also a good estimation for the damping of the motion even when $v_0$ approaches $v_{max}.$ This is evident from (\ref{vsqt}) where time $t$ comes with a factor of $\beta=\fr{s^2}{1+s^2}$.

\begin{figure}
	\centering
		 \epsfig{file=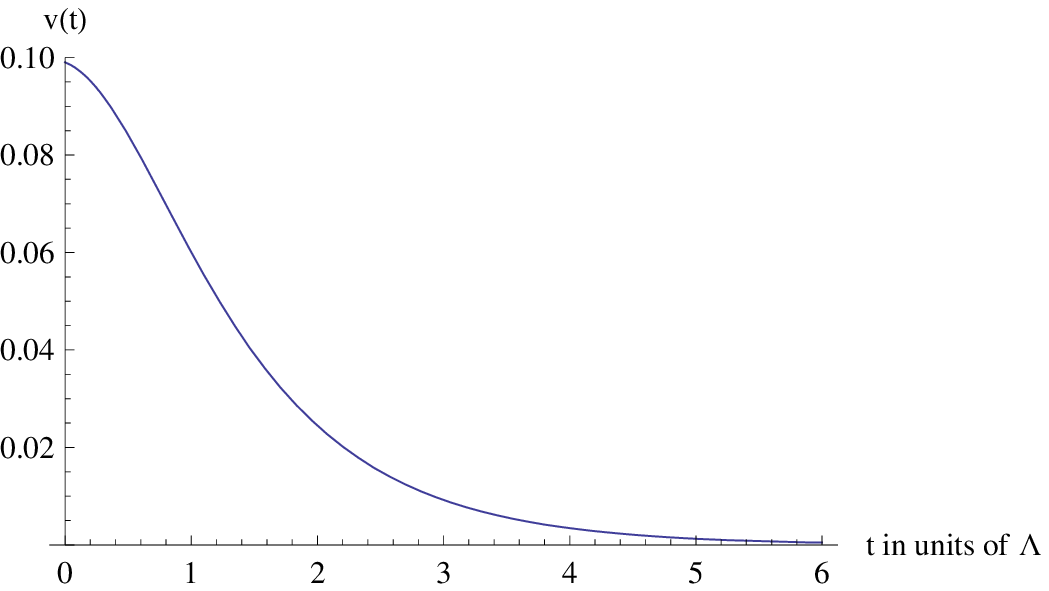,height=2in,width=3in}\epsfig{file=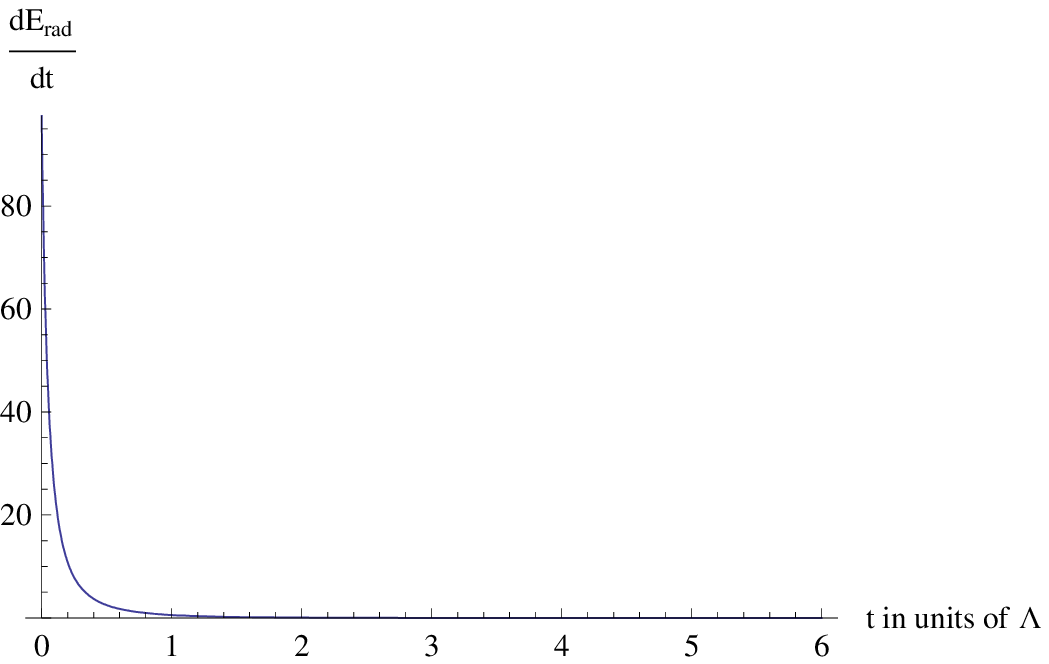,height=2in,width=3in}
		\caption{\textbf{Left}:Velocity of the particle for large $s=10$ and initial velocity $v_0=0.099c\approx v_{max}.$ \qquad
		\textbf{Right}: Rate of energy transfer from the particle per unit mass for large $s=10$ and initial velocity $v_0=0.099c\approx v_{max}.$}
	\label{fig:velrells}
\end{figure}

Therefore, in the case of small $s=0.1,$ we expect a timescale for the damping of the velocity $t_{damp}=\beta^{-1}\approx 100.$ In the case of large $s=10,$ we expect a timescale $t_{damp}=\beta^{-1}\approx 1.$
This is indeed observed in the figures \ref{fig:velnonrelss} , \ref{fig:velrelss} for small $s=0.1$ where the timescale for the damping is $t_{damp}\approx 100$ and for large $s=10$ in figures \ref{fig:velnonrells}, \ref{fig:velrells}
where indeed $t_{damp}\approx 1.$

Because the motion is damped, the velocity of the quark will be always decreasing.For late times $t>>\beta^{-1}=\fr{1+s^2}{s^2},$ it will have an exponential form because when the velocity $\vec{v}(t)$ of the quark  obeys $||\vec{v}(t)||<<v_{max}=\fr{1}{\sqrt{1+s^2}},$ the linear approximation (\ref{linsol}) is valid giving an exponential form for the velocity.
In the case that the initial velocity satisfies this criterion ($v_0<<v_{max}=\fr{1}{\sqrt{1+s^2}}$), the velocity as a function of time is of exponential form as is seen in the left figure \ref{fig:velnonrelss} for small $s=0.1$, and in the left figure \ref{fig:velnonrells} for large $s=10$.
The corrections to the linear approximation are of order $v_0^2$ and $s^2v_0^2.$

Therefore, for small $s$, the relevant next order correction to the linear approximation is of order $v_0^2$ and therefore for $s=0.1,v_0=0.1c<<v_{max}\approx c$ we expect the relative difference of linear and the numerical solution to be $\approx 1\%$ in accordance with the value $0.25\%$ observed in the right figure \ref{fig:velnonrelss}.\\
On the other hand for large $s,$ the relevant next order correction to the linear approximation is of order $s^2v_0^2$ and therefore for $s=10,v_0=0.01c<<v_{max}\approx 0.1c$ we expect the relative difference of linear and the numerical solution to be $\approx 1\%$ in accordance with the value $0.25\%$ observed in the right figure \ref{fig:velnonrells}.\\
\indent

We therefore confirm that the linear approximation is valid for the range of velocities $||\vec{v}(t)||<<v_{max}.$
Furthermore, even in the case where $v_0\approx v_{max},$ the endpoint will decelerate and therefore the linear approximation will be be valid for large times $t>>\beta^{-1}=\fr{1+s^2}{s^2}$.
In the left figure \ref{fig:velrelss} we show the case for a motion with $v_0\approx v_{max}$ for small parameter $s,$ and in the left figure \ref{fig:velrells} for large parameter $s.$ For large times $t>>\beta^{-1}=\fr{1+s^2}{s^2},$ the velocity has a clear exponential form. However, for small times with respect to $\beta=\fr{s^2}{1+s^2},$ the velocity has negative second derivative which later turns into a positive one as expected for an exponential form.

\subsection{The induced world-sheet geometry.}

We consider as worldsheet coordinates the proper time $\tilde{\tau}$ on the auxiliary boundary at $r=0$ and $r$ the radial coordinate of AdS with radius $L$.\\
Then the induced metric elements with respect to the worldsheet coordinates $\tilde{\tau},r$ become
\be
g_{\tilde{\tau}\tilde{\tau}}=-\fr{L^2}{r^2}\lt(1-r^2\ddot{\tilde{x}}^{\mu}(\tilde{\tau})\ddot{\tilde{x}}_{\mu}(\tilde{\tau})\rt),g_{\tilde{\tau} r}=-\fr{L^2}{r^2}\sp g_{rr}=0
\ee
where $\tilde{x}^{\mu}=\{\tilde{t},\tilde{x},\tilde{y},\tilde{z}\}$ and $\tilde{\tau}$ is the proper time on the boundary at $r=0.$
In appendix \ref{wshor} we study the existence of a world-sheet horizon and its position.
We have done this in the linear approximation $v_0<<v_{max}.$
The result is that the induced metric has a horizon which moves towards $r=\infty$ and for late times it moves exponentially fast with time as in (\ref{horasym}).

By calculating the curvature invariant we have a constant curvature $R=-\fr{2}{L^2}$ everywhere and for all times, the curvature of a hyperboloid with radius $L$ in two dimensions.
This constant negative curvature is true for every solution of Mikhailov \cite{Mikhailov's}\footnote{We remind that by tilded quantities we name the coordinates at the auxiliary boundary at $r=0$ with which the solution of Mikhailov has the simplest form.}
\be
X^\mu(\tilde{\tau},r)=\tilde{x}^\mu(\tilde{\tau})+r\fr{d\tilde{x}^\mu(\tilde{\tau})}{d\tilde{\tau}}.
\ee

The following remarks are relevant:

\begin{enumerate}
	\item The position of the horizon is time-dependent due to the fact that the motion of the quark is damped.
 \item Because the metric elements and the position of the horizon are time-dependent the definition of a Hawking temperature in this problem is not straightforward.
 \item The profile of the string $\vec{X}(X^0,r)$ is problematic because the function $X^0(t,r)$ is not invertible for all $r$ (1-1 function between $t$ and $X^0$) in order to have $t(X^0,r)$ and substitute it in the expressions $\vec{X}(t,r)$ in (\ref{stringsol}).
  In our case with a constant magnetic field $B$ and the convention $\Lambda=1,$ equations (\ref{stringsol}) become\footnote{We remind that $\hat{z}$ is the vector with norm $1$ in the $z-$direction.}
\bea
X^0(t,r)&=&\fr{r-1}{\sqrt{1-\fr{s^2\vec{v}(t)^2}{1-\vec{v}(t)^2}}}\fr{1}{\sqrt{1-\vec{v}(t)^2}}+t \label{x0tr} \\
\vec{X}(t,r)&=&\fr{r-1}{\sqrt{1-\fr{s^2\vec{v}(t)^2}{1-\vec{v}(t)^2}}}\fr{1}{\sqrt{1-\vec{v}(t)^2}}\lt(\vec{v}(t)-s\lt(\hat{z}\times\vec{v}(t)\rt)\rt)+\vec{x}(t) \;.\label{xitr}
\eea
The minimum of $X^0(t,r)$ is at
\be
\fr{\pt X^0}{\pt t}=0\Rightarrow r^{\ast}(t)=1-\frac{\left(1-\left(s^2+1\right) \vec{v}(t)^2\right)^{3/2}}{\left(s^2+1\right) \vec{v}(t) \cdot\vec{a}(t)}=1-\frac{\left(1-\fr{ \vec{v}(t)^2}{v_{max}^2}\right)^{3/2}}{\fr{\vec{v}(t) \cdot\vec{a}(t)}{v_{max}^2}}=
\ee
\be
=1+\frac{ \sinh \left(2 \left(\beta\ t+\tanh^{-1}\left(\sqrt{1-\fr{v_0^2}{v_{max}^2}}\right)\right)\right) \tanh \left(\beta\ t+\tanh ^{-1}\left(\sqrt{1-\fr{v_0^2}{v_{max}^2}}\right)\right)}{2 \beta},\nonumber \\
\label{rastt}
\ee
and it is an increasing function of $t.$
We show the function $X^0(t,r)$ for different $r$ in figure \ref{fig:x0t}.

\begin{figure}
	\centering
		 \epsfig{file=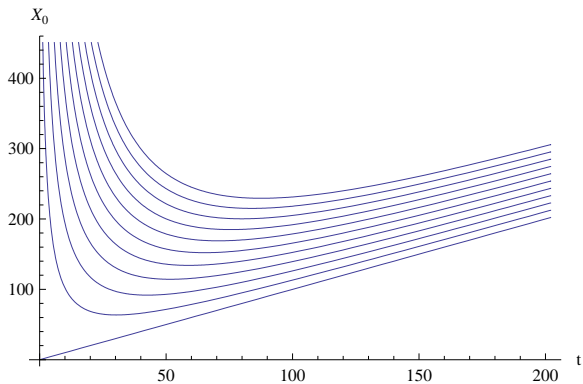,height=2in,width=3in}
		\caption{The function $X^0(t,r)$ for $r=1,11,21,...$ with s=0.1 and $v_0^2=0.9999v_{max}^2.$We measure $X^0,t,r$ in units of $\Lambda.$}
	\label{fig:x0t}
\end{figure}

We choose to invert $X^0(t,r)$ by selecting the right branch (i.e. for $t>t_{min}$, where $t_{min}$ is the time 
$t$ of the minimum of $X^0(t,r)$) in order for $X^0$ to increase as we increase $t.$
For a fixed time $X^0=T$ we can invert $X^0(t,r)$ and draw the string up to a point $r_{max}(T).$ This is given by the equation $T=X^0(t(r_{max}),r_{max})$ where $t(r_{max})$ is the inversion of $r^{\ast}(t)$ in (\ref{rastt}).

\end{enumerate}

\subsection{Profile of the string}

We have calculated numerically the motion of the string from equation (\ref{mikhsol}) which gives us $X^{\mu}(\tau,r)$ and from $\fr{dt}{d\tau}=\fr{1}{\sqrt{1-\vec{v}^2(t)}}$ we can find the functions $X^{\mu}(t,r).$

\begin{figure}[ht!]
	\centering
		 \epsfig{file=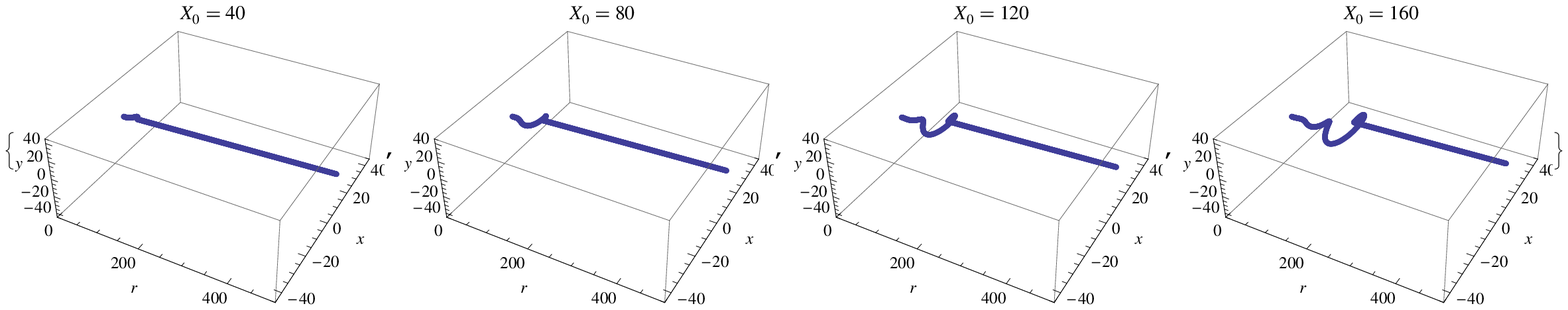,height=1.4in,width=6.7in}\\
		 \epsfig{file=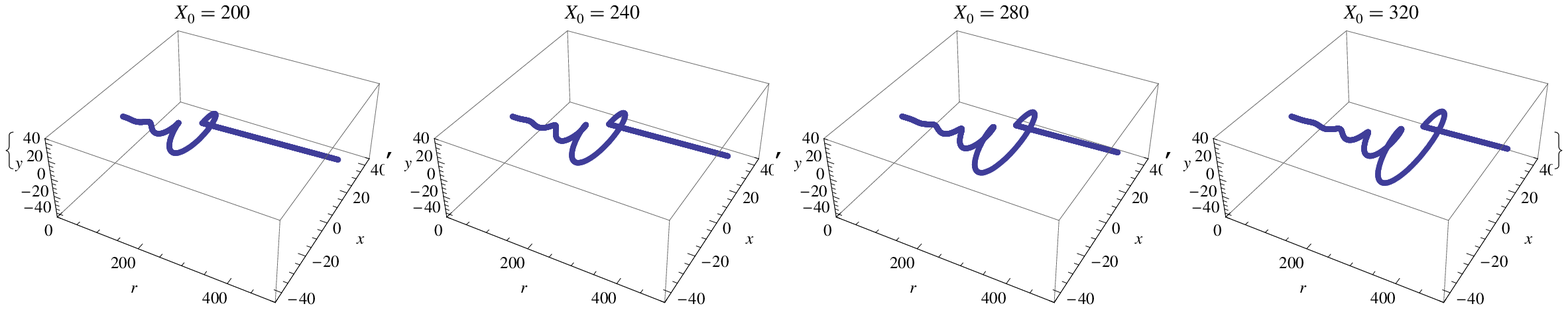,height=1.4in,width=6.7in}\\
		 \epsfig{file=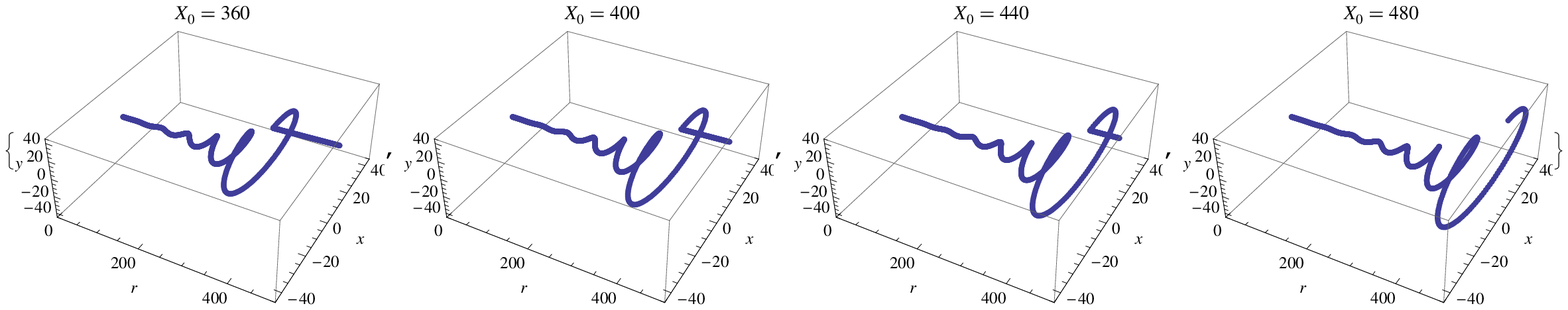,height=1.4in,width=6.7in}
		\caption{Profile of the string in the (more interesting) case of small $s=0.1$ and initial velocity $v_0=0.1v_{max}\approx0.1c$ for different times $X^0$ in the range $r\in(1,400)$ for the radial coordinate.We measure $X^0,r,x,y$ in units if $\Lambda.$}
	\label{fig:stringprof}
\end{figure}

Then, we invert the function $X^0(t,r)$ to $t(X^0,r)$ and we substitute $t$ in the functions $\vec{X}\lt(t,r\rt)$ in order to find $\vec{X}\lt(X^0,r\rt)$ which is necessary in order to draw the string.
In figure \ref{fig:stringprof} we show the profile of the string for different times $X^0.$

Furthermore, we consider the endpoint (quark) to move with constant velocity $v_0$ for times $t\in(-\infty,0)$ and in a constant magnetic field for times $t\in(0,\infty).$ We consider it moving for $t<0$ in the $y$-direction and then we have the damped circular motion for $t>0.$ In our figure we consider small $s=0.1$ and initial velocity $v_0=0.1\ v_{max}.$

We clearly observe the spiral motion of the boundary to propagate down the string and become bigger with distance. We also clearly see the abrupt change of its behaviour (discontinuity in its derivative) at the point where the propagation of the motion at constant velocity and in constant magnetic field meet.

Finally, we also observe the motion of the string to be damped at each $r$, because the boundary motion is also damped.

\section{Hall effect at zero temperature \label{halleff}}

To study the analogue of the Hall effect we must assume a small electric field $E_x$ in the x-direction and a large magnetic field $B_z$ in the z-direction.
We describe the motion of the charge carriers in a strongly coupled vacuum at zero temperature with the motion of the endpoint of the strings at $r=\Lambda,$ where $\Lambda^{-1}$ is proportional to the mass of the carriers.
Pure AdS space-time has Lorentz invariance under boosts and rotations in the $x,y,z,t$ directions. The electric and magnetic field transform under boosts:

\be
\vec{E}_{||}=\vec{E}_{||}\sp
\vec{E}_{\bot}=\gamma\lt(\vec{E}_{\bot}+\vec{v}\times\vec{B}\rt)\sp
\vec{B}_{||}=\vec{B}_{||},\\
\vec{B}_{\bot}=\gamma\lt(\vec{B}_{\bot}-\vec{v}\times\vec{E}\rt)
\ee
By doing a boost with velocity $v_y=-E_x/B_z$ we have in the boosted frame
\be
\vec{E}'=\vec{0}\sp
\vec{B}'=\hat{z} \sqrt{B_z^2-E_x^2}
\ee
Therefore in this frame we have only a constant magnetic field $B_z'$ in the z-direction and the motion of the string will be a spiral towards a fixed point as we have seen already.
Therefore for large times,  the velocity of the particle will be that of the boosted frame $v_y=-E_x/B_z.$
From this,  we deduce that we have the Hall conductivity
\be
\sigma_{xy}=\fr{j_y}{E_x}=-\fr{q}{B_z}.
\ee

\section{Electromagnetic fields needed in order to have a circular motion for the quark}

As we have seen in the case of a constant magnetic field on a brane in pure AdS, the motion of the string is damped and after some time it will stop practically moving. However in \cite{MITgroup},\cite{Fadafan},\cite{ChernicoffParedes}, the case of a circular motion of the string has been discussed.\\
Therefore it is interesting to examine what kind of electromagnetic fields on the boundary are necessary in order to satisfy the boundary conditions for the aforementioned circular motion.\\
We assume a constant magnetic field $\vec{B}=B\hat{z},$ and a time-dependent electric field $\vec{E}(t)$ on the $x-y$ plane.
For convenience we use $\vec{\mathcal{E}}$ defined by $\vec{E}=\fr{m}{\Lambda}\vec{\mathcal{E}}.$
We also choose the unit system where $\Lambda=1$.
We need for the electric field a component codirectional with the velocity of the particle in order to compensate for the drag force that decelerates it, and this is enough, as the magnetic field induces a force perpendicular to the velocity of the particle.\\
Therefore we assume the form $\vec{\mathcal{E}}=\mathcal{E}_0\vec{v}(t)$ for the electric field, i.e. a time dependent electric field (rotating) that is always codirectional with the (rotating) velocity of the particle with $\mathcal{E}_0$ being the factor of proportionality.
Then the equation (\ref{eoms}) gives instead of (\ref{eomst})
\bea
\fr{d}{dt}\lt(\gamma\fr{\fr{d\vec{x}}{dt}-s\fr{d\vec{x}}{dt}\times\hat{z}-m \mathcal{E}_0\fr{d\vec{x}}{dt} }{\sqrt{1-s^2\gamma^2\lt(\fr{d\vec{x}}{dt}\rt)^2}}\rt)&=&\fr{s\fr{d\vec{x}}{dt}\times\hat{z}+m \mathcal{E}_0\fr{d\vec{x}}{dt}-s^2\gamma^2\lt(\fr{d\vec{x}}{dt}\rt)^2\fr{d\vec{x}}{dt}}{1-s^2\gamma^2\lt(\fr{d\vec{x}}{dt}\rt)^2}
\label{eomstn}
\eea
and when we substitute the ansatz motion $x(t)=R\cos(\omega t),y(t)=R\sin(\omega t),$ in order for the equations (\ref{eomstn}) to be satisfied we must have
\bea
\mathcal{E}_0=\frac{\left(s^2-1\right) \vec{v}^2+\sqrt{4 s^2 \left(\vec{v}^2-1\right)+\left(\left(s^2-1\right) \vec{v}^2+1\right)^2}+1}{2 \left(1-\vec{v}^2\right)},
\eea
and the angular velocity $\omega$ is given by
\be
\omega =\frac{s \sqrt{1-\vec{v}^2} \sqrt{\frac{\left(s^2+1\right) \vec{v}^2-1}{\vec{v}^2-1}}}{(\mathcal{E}_0-1)^2+s^2}
\ee
with respect to $s=\fr{B \Lambda}{m},$ i.e.  the magnetic field strength and the radius is found by
\be
R=\fr{||\vec{v}||}{\omega}
\ee
as the norm of the velocity is constant during the circular motion.

\vskip 3cm
\addcontentsline{toc}{section}{Acknowledgements}
\section{Acknowledgements}\label{ACKNOWL}

We would like to thank Costas Bachas and Tassos Taliotis for useful conversations. 
We would also like to thank the referee for his very constructive suggestions. This work was partially supported by
a European Union grant FP7-REGPOT-2008-1-CreteHEP Cosmo-228644, and PERG07-GA-2010-268246.

\newpage
 \addcontentsline{toc}{section}{Appendices}
\appendix
\section*{Appendix}
\section{The world-sheet horizon} \label{wshor}

\begin{figure}[ht!]
	\centering
		\epsfig{file=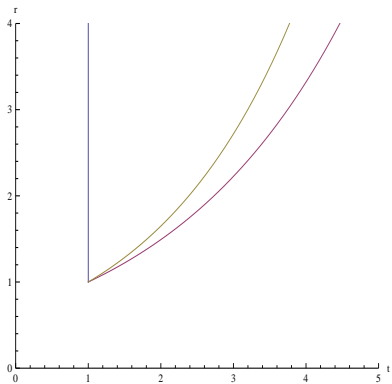,height=2in,width=2in}
		\caption{Two null rays and a timelike ray that begin from $t_0=1,r_0=1$.The timelike ray lies between the two null rays.}
	\label{fig:ntn}
\end{figure}

In this section we will find  the world-sheet horizon in the approximation where the initial velocity $v_0$ of the endpoint at $r=\Lambda$ is very small $v_0<<v_{max}=\fr{1}{\sqrt{1+s^2}}.$ We have to examine the null trajectories starting from each space-time point because the timelike ones will lie between the null ones (e.g. figure \ref{fig:ntn}).

The background AdS metric is
\bea
ds^2=-\fr{L^2}{r^2}\lt(-dt^2+dr^2+dx^2+dy^2+dz^2\rt),
\eea
and the induced metric of the world-sheet is
\bea
ds^2=-\fr{L^2}{r^2}\lt(1-r^2\ddot{\tilde{x}}^\mu \ddot{\tilde{x}}_\mu\rt)d\tilde{\tau}^2-\fr{L^2}{r^2}d\tilde{\tau}\ dr
\eea
where $\tilde{x}^\mu,\tilde{\tau}$ are the coordinates and the proper time of the auxiliary boundary at $r=0$ and by $\ddot{\tilde{x}}^\mu$ we mean $\fr{d^2\tilde{x}^\mu}{d\tilde{\tau}^2}.$

\begin{figure}[ht!]
	\centering
		\epsfig{file=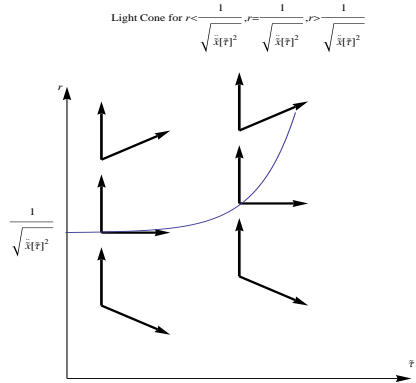,height=2in,width=2.5in}
		\caption{The light cone for distinct cases of $r$.}
	\label{fig:horizondyn}
\end{figure}

The light cone at each spacetime point $(\tilde{\tau},r)$ is formed by the light rays that obey
\bea
ds^2&=&-\fr{L^2}{r^2}\lt(1-r^2\ddot{\tilde{x}}^\mu(\tilde{\tau})\ddot{\tilde{x}}_\mu(\tilde{\tau})\rt)d\tilde{\tau}^2-\fr{L^2}{r^2}d\tilde{\tau} dr=0\Rightarrow\nonumber \\
d\tilde{\tau}&=&0\;\textit{or}\;\fr{dr}{d\tilde{\tau}}=-\lt(1-r^2\  \ddot{\tilde{x}}^\mu(\tilde{\tau})\ddot{\tilde{x}}_\mu(\tilde{\tau})\rt)
\label{lcdiffeq}
\eea
as shown in figure \ref{fig:horizondyn} for the three cases $r<\fr{1}{\sqrt{\ddot{\tilde{x}}^2}},r=\fr{1}{\sqrt{\ddot{\tilde{x}}^2}},r>\fr{1}{\sqrt{\ddot{\tilde{x}}^2}}$.

\indent

\paragraph{The static case}

\begin{figure}[ht!]
	\centering
		\epsfig{file=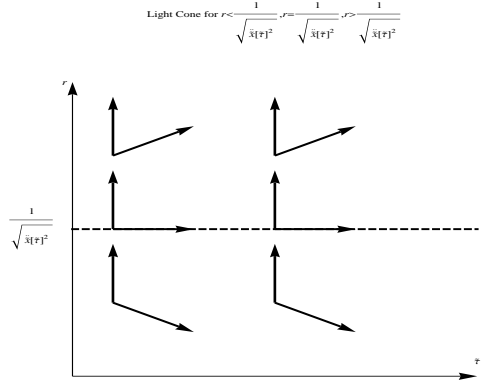,height=2in,width=2.5in}
		\caption{The light cone for distinct cases of $r$.}
	\label{fig:horizonstatic}
\end{figure}

In the time-independent case where $\ddot{\tilde{x}}(\tilde{\tau})^2=const,$ the two light rays that begin from points at $r>\fr{1}{\sqrt{\ddot{\tilde{x}}^2}},$ are directed towards $r=\infty$ and no light rays can reach the points with $r<\fr{1}{\sqrt{\ddot{\tilde{x}}^2}}.$
This is easily seen in figure \ref{fig:horizonstatic} where the light cone is directed towards $r=\infty.$
Therefore, the part $r>\fr{1}{\sqrt{\ddot{\tilde{x}}^2}}$ is causally disconnected from the part $r<\fr{1}{\sqrt{\ddot{\tilde{x}}^2}},$ and the point $r=\fr{1}{\sqrt{\ddot{\tilde{x}}^2}}$ is a horizon.

\indent

\paragraph{The dynamical case}

On the other hand, in the time-dependent metric of our model, the function $\fr{1}{\sqrt{\ddot{\tilde{x}}^2}}$ increases with time and in the linear approximation it moves towards $r=\infty$ exponentially fast with time and the differential equation in (\ref{lcdiffeq}) for the one light ray is
\bea
\fr{dr}{d\tilde{\tau}}=-\lt(1-r^2\ \ddot{\tilde{x}}^\mu(\tilde{\tau})\ddot{\tilde{x}}_\mu(\tilde{\tau})\rt).
\eea
The coordinates $x^{\mu}$ and the proper time $\tau$ of the boundary at $r=\Lambda=1$ are related
to the coordinates $\tilde{x}^\mu$ and proper time $\tilde{\tau}$ of the auxiliary boundary at $r=0$ by \cite{ChernicoffGuijosa}:
\bea
d\tau&=&d\tilde{\tau} \sqrt{1-\ddot{\tilde{x}}^2}\sp \ddot{\tilde{x}}^2=\fr{4\pi}{\lambda}F_{\nu\mu}F^{\rho\mu}\fr{dx^{\nu}}{d\tau}\fr{dx_{\rho}}{d\tau}=\fr{4\pi}{\lambda}\fr{B^2v(t)^2}{1-v(t)^2}=\fr{s^2v(t)^2}{1-v(t)^2},
\label{traux}
\eea
where $\vec{v}(t)$ is the velocity of the endpoint at $r=\Lambda=1.$
We remind the reader the conventions $s=\fr{B}{m}$ for the magnetic field strength w.r.t. to the mass of the particle $m$ and $\Lambda=\fr{2\pi m}{\sqrt{\lambda}}=1$.By choosing $\Lambda=1,$ we actually measure $\tilde{x}^\mu,\tilde{\tau},x^\mu,\tau$ in units of $\Lambda.$

Then, using (\ref{traux}), the differential equations (\ref{lcdiffeq}) for the light-rays become
\bea
dt&=&0\; \;\;{\rm or}\;\; \; \fr{dr}{dt}\fr{dt}{d\tau}\fr{d\tau}{d\tilde{\tau}}=-\lt(1-r^2\fr{s^2v(t)^2}{1-v(t)^2}\rt)\Rightarrow\nonumber \\
\fr{dr}{dt}&=&-\lt(1-r^2\fr{s^2v(t)^2}{1-v(t)^2}\rt)\sqrt{\fr{1-v(t)^2}{1-\fr{s^2v(t)^2}{1-v(t)^2}}}.
\label{lcdiffeq2}
\eea
The evolution of the norm of velocity of the endpoint can be found analytically and was given in (\ref{vsqt}) that we reproduce here
\bea
v(t)=\fr{sech\lt(\fr{s^2}{1+s^2}t+tanh^{-1}\lt(\sqrt{1-\lt(1+s^2\rt)v_0^2}\rt)\rt)}{\sqrt{1+s^2}},
\eea
where $v_0$ is the initial velocity of the endpoint at $r=\Lambda=1$.
The second differential equation in (\ref{lcdiffeq2})  becomes

\bea
\fr{dr}{dt}&=&\frac{\coth \left(\frac{s^2 t}{s^2+1}+\tanh ^{-1}\left(\sqrt{1-\left(s^2+1\right) v_0^2}\right)\right) }{s^2+1}\times\nonumber \\
&\times&\left(\left(r^2 s^2+1\right) sech^2\left(\frac{s^2 t}{s^2+1}+\tanh ^{-1}\left(\sqrt{1-\left(s^2+1\right)
   v_0^2}\right)\right)-s^2-1\right).
\label{lcdiffeq3}
\eea
To solve it  we will assume that the endpoint at $r=\Lambda=1$ has at $t=0$ very small initial velocity $v_0$ (i.e. $v_0<<\fr{1}{\sqrt{1+s^2}}$) so that
\bea
\coth \left(\frac{s^2 t}{s^2+1}+\tanh ^{-1}\left(\sqrt{1-\left(s^2+1\right) v_0^2}\right)\right)&\approx&1\nonumber\\
sech\left(\frac{s^2 t}{s^2+1}+\tanh ^{-1}\left(\sqrt{1-\left(s^2+1\right) v_0^2}\right)\right)&\approx&a_0e^{-\fr{s^2t}{1+s^2}}
\label{linapprox}
\eea
with $a_0=\fr{2}{e^{\tanh^{-1}\lt(\sqrt{1-(1+s^2)v_0^2}\rt)}}.$
Therefore, in the case where the initial velocity is much smaller than the maximum possible one (i.e. $v_0<<v_{max}=\fr{1}{\sqrt{1+s^2}}$), then $a_0\approx v_0\sqrt{1+s^2}$ and (\ref{lcdiffeq3}) becomes
\bea
\fr{dr}{dt}=a_0^2 \fr{s^2}{1+s^2} r(t)^2 e^{-\frac{2 s^2 t}{s^2+1}}-1=-(1-r(t)^2A^2e^{-2\beta\ t})
\label{lcdiffeq4}
\eea
with $\beta=\fr{s^2}{1+s^2}$ and $A=\fr{s}{\sqrt{1+s^2}} a_0=s v_0.$
Then (\ref{lcdiffeq4}) with initial condition $r(0)=r_0$ gives as solution
\bea
r(t;r_0)=\frac{e^{\beta  t} \left(\left(A r_0 I_1\left(\frac{A}{\beta }\right)-I_0\left(\frac{A}{\beta }\right)\right) K_0\left(\frac{A e^{-t \beta }}{\beta }\right)+\left(K_0\left(\frac{A}{\beta }\right)+A r_0 K_1\left(\frac{A}{\beta}\right)\right) I_0\left(\frac{A e^{-t \beta }}{\beta }\right)\right)}{A \left(\left(K_0\left(\frac{A}{\beta }\right)+A r_0 K_1\left(\frac{A}{\beta }\right)\right) I_1\left(\frac{A e^{-t \beta }}{\beta }\right)+\left(I_0\left(\frac{A}{\beta}\right)-A r_0 I_1\left(\frac{A}{\beta }\right)\right) K_1\left(\frac{A e^{-t \beta }}{\beta }\right)\right)}.
\label{rtr0}\nonumber \\
\eea
where $I_n,K_n$ are the modified Bessel functions of the first kind\footnote{The quantity $\fr{A}{\beta}=\fr{\lt(1+s^2\rt)\ v_0}{s}=\fr{v_0}{v_{max}\sqrt{\beta}}$ can't be considered small because $\beta\in(0,1)$ and for small $\beta$ the quantity $\fr{v_0}{v_{max}\sqrt{\beta}}$ may not be small.}.

In order to find the horizon for each time $t,$ we need to examine all the timelike and lightlike trajectories beginning from any point of our spacetime $r,t.$ This can be done by examining only the light rays because all the timelike rays that start from the same point will lie between the two light-rays for each $r$,i.e. $t_{null_1}(r)>t_{timelike}(r)>t_{null_2}(r)$ as seen in figure \ref{fig:ntn}. One of the light-rays obeys $dt=0$ and therefore has constant $t_{null_1}(r)=t_0$ and the other light-ray obeys the differential equation (\ref{lcdiffeq4}) which gives a solution $t_{null_2}(r)$.
Furthermore, we may solve the equation for the second light ray (\ref{lcdiffeq4}) only for initial time $t=0$ and initial position $r_0,$ because each light-ray that obeys the same differential equation and begins from another spacetime point (e.g. $t_0>0,r_0'$)  will be an extension to a larger time of a light ray beginning from $t=0$ and their behaviour for large parameter $t$ will be common.

\begin{figure}[ht!]
	\centering
		\epsfig{file=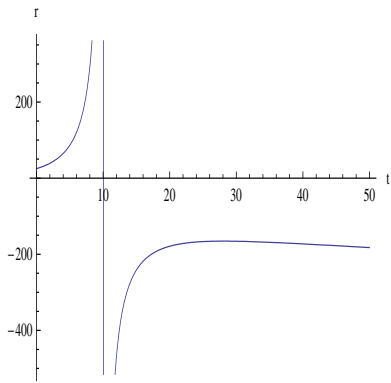,height=2in,width=2in}
		\caption{The light ray starts from $r_0=25$ and diverges to $r=\infty$ at finite $t\approx10$.}
	\label{fig:lrdiv}
\end{figure}

\begin{figure}[ht!]
	\centering
		\epsfig{file=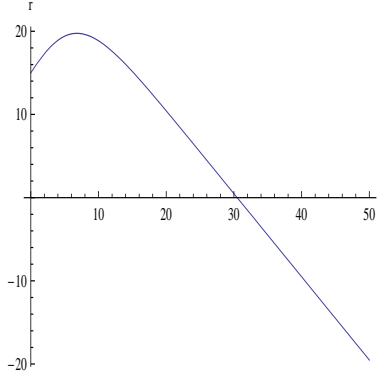,height=2in,width=2in}
		\caption{The light ray starts from $r_0=15$ and turns back towards $r=0$.}
	\label{fig:lrconv}
\end{figure}

The solution $r(t;r_0)$ in (\ref{rtr0}) for some values $r_0$ of the initial position of the light ray (above a bound $r_{hor}(0)$) goes to $r=\infty$ at finite $t$ as in figure \ref{fig:lrdiv}, while for some others (below the bound $r_{hor}(0),$ i.e. $r_0<r_{hor}(0)$) it returns back towards $r=0$ (figure \ref{fig:lrconv}).

In both cases at $t=\infty$ the function $r(t;r_0)$ has derivative $-1.$\footnote{In the case there is a divergence of $r(t,r_0)$ at finite $t$ as in figure \ref{fig:lrdiv}, the solution to the differential equation (\ref{lcdiffeq4}) can be extended to $t$ higher than $t_{inf}$ where the infinity is reached. The trajectory at $t>t_{inf}$ describes light-rays that would start from $t>t_{inf}$ at points $\lt(t,r\lt(t;r_0\rt)\rt)$ from which the trajectory $r(t;r_0)$ passes and as we see in figure \ref{fig:lrdiv} they would turn back towards $r=0.$ }
This can be seen by the asymptotic expansion of $r(t;r_0)$ for large $t$
\bea
r(t\rightarrow\infty;r_0)=\frac{\log \left(\frac{A}{2 \beta }\right)+\frac{\gamma  I_0\left(\frac{A}{\beta }\right)+K_0\left(\frac{A}{\beta }\right)+A r_0 \left(K_1\left(\frac{A}{\beta }\right)-\gamma  I_1\left(\frac{A}{\beta }\right)\right)}{I_0\left(\frac{A}{\beta }\right)-A r_0 I_1\left(\frac{A}{\beta }\right)}}{\beta }-t.
\eea
Therefore, the case that seperates the two behaviours is the case where the function $r(t;r_0)$ diverges to $r=\infty$ at  $t=\infty$ and then the function $r(t;r_0)$ with the suitable $r_0$ is the function $r_{hor}(t),$ the position of the horizon at each time $t.$
The case where $r(t;r_0)$ diverges at $t\rightarrow\infty$ is when the constant
$$\frac{\log \left(\frac{A}{2 \beta }\right)+\frac{\gamma  I_0\left(\frac{A}{\beta }\right)+K_0\left(\frac{A}{\beta }\right)+A r_0 \left(K_1\left(\frac{A}{\beta }\right)-\gamma  I_1\left(\frac{A}{\beta }\right)\right)}{I_0\left(\frac{A}{\beta }\right)-A r_0 I_1\left(\frac{A}{\beta }\right)}}{\beta }$$
becomes infinite at $t=\infty,$ and this doesn't allow for a divergence before $t=\infty$ and obviously not a turnback of $r(t;r_0).$ This happens when the constant $r_0$ is
$$r_0=\frac{I_0\left(\frac{A}{\beta}\right)}{A I_1\left(\frac{A}{\beta}\right)}$$
and then $r(t;r_0)$ in (\ref{rtr0}) becomes $r_{hor}(t)$ which is:
\bea
r_{hor}(t)=\frac{e^{\beta t} I_0\left(\frac{A e^{-\beta t}}{\beta}\right)}{A I_1\left(\frac{A e^{-\beta t}}{\beta}\right)},
\label{rhorfin}
\eea
with $A=s\ v_0,\beta=\fr{s^2}{1+s^2}=1-v_{max}^2.$\footnote{By $v_{max}=\fr{1}{\sqrt{1+s^2}}$ we name the maximum possible initial velocity.}
Any timelike or lightlike trajectory that starts from the point $(t_0,r_0)$ with $r_0>r_{hor}(t_0)$ will reach $r=\infty$ at finite retarded time $t.$
On the other hand, one null\footnote{It is the light-ray that obeys $t=const.$} and some timelike trajectories starting from $(t_0,r_0)$ with $r_0<r_{hor}(t_0)$ will reach $r=\infty$ at finite retarded time $t,$ whereas the other null ray and other timelike trajectories will return back towards the boundary $r=\Lambda.$
Therefore, the function $r_{hor}(t)$ in (\ref{rhorfin}) describes the position of the horizon with time.

The function $r_{hor}(t)$ behaves for large times $t\rightarrow\infty$ as
\bea
r_{hor}(t)\rightarrow\fr{2\beta}{A^2}e^{2\beta t}=\fr{2v_{max}^2}{v_0^2}e^{2\beta t},
\label{horasym}
\eea
with $\beta=\fr{s^2}{1+s^2},$ i.e. the horizon moves from the boundary at $r=\Lambda$ exponentially fast with time for large times.

\indent

We can also check under which conditions our approximation (\ref{lcdiffeq4}) is valid.
By expanding (\ref{lcdiffeq2}) to next order in $v_0$ we have the differential equation

\bea
\fr{dr}{dt}=-1+\frac{1}{2} v_0^2 \left(1-s^2\right) e^{-2 \beta  t}+\frac{1}{2} r^2 s^2 v_0^2 \left(v_0^2\lt(1+s^2\rt)+2\right) e^{-2 \beta  t}
\eea
which is of the form

\bea
\fr{dr}{dt}=-1+B\ e^{-2\beta t}-r(t)^2A'^2e^{-2\beta t}
\eea
with constants $B=\frac{1}{2} \left(1-s^2\right) v_0^2\sp A'=\sqrt{\frac{1}{2} \left(s^2+1\right) v_0^4 s^2+v_0^2 s^2}.$
It can be written

\bea
\fr{dr}{dt}+1-r(t)^2A^2e^{-2\beta\ t}=B\ e^{-2\beta t}-\lt(A'^2-A^2\rt)r(t)^2e^{-2\beta t}
\eea

where $|B|=\frac{1}{2} \left|1-s^2\right| v_0^2<\fr{1}{2}\lt(1+s^2\rt)v_0^2=\fr{1}{2}\fr{v_0^2}{v_{max^2}}$ and $A'^2-A^2=\frac{1}{2} \left(s^2+1\right) v_0^4 s^2=\fr{v_0^2}{2v_{max^2}}s^2v_0^2.$
This means that the factors of the right hand side are suppressed by $\fr{v_0^2}{v_{max}^2}$ with respect to the corresponding factors of the left-hand side

\bea
|B|<\fr{1}{2}\fr{v_0^2}{v_{max^2}}<<1\sp A'^2-A^2=\frac{1}{2} \fr{v_0^2}{v_{max}^2}s^2v_0^2<<s^2v_0^2=A^2.
\eea

Therefore, we can rely on the approximate solution (\ref{rhorfin}) for small $\fr{v_0}{v_{max}}\ll1$ with relative error $\fr{\delta r_{hor}(t)}{r_{hor(t)}}\sim\fr{v_0^2}{v_{max}^2}.$

\end{document}